\documentclass[12pt,aps,prb,reprint,superscriptaddress,showpacs,floatfix]{revtex4-1}
\usepackage{graphicx}
\usepackage{amssymb}
\usepackage{color}
\usepackage{bm}
\usepackage[english]{babel}
\usepackage[a4paper=true,pagebackref=false]{hyperref}
\hypersetup{colorlinks,citecolor=blue,filecolor=blue,linkcolor=blue,urlcolor=blue}
\hypersetup{pdftitle={The Mn site in Mn-doped Ga-As nanowires: an EXAFS study}}

\begin{document}

\title{The Mn site in Mn-doped Ga-As nanowires: an EXAFS study}

\author{F. d'Acapito}
\affiliation{CNR-IOM-OGG c/o ESRF GILDA CRG BP220 F-38043 Grenoble (France)}

\author{M. Rovezzi}
\affiliation{European Synchrotron Radiation Facility BP220 F-38043 Grenoble (France)}

\author{F. Boscherini}
\affiliation{Department of Physics, University of Bologna, Viale Berti Pichat 6/2, I-40127 Bologna Italy}
\affiliation{CNR-IOM-OGG c/o ESRF GILDA CRG BP220 F-38043 Grenoble (France)}

\author{F. Jabeen}
\affiliation{CNR-IOM, SS14 Km 163.5, I-34149 Basovizza Trieste, Italy }
\affiliation{Sincrotrone Trieste S.C.p.A., SS14 Km 163.5, I-34149 Basovizza Trieste, Italy}

\author{G. Bais}
\affiliation{CNR-IOM, SS14 Km 163.5, I-34149 Basovizza Trieste, Italy }
\affiliation{Sincrotrone Trieste S.C.p.A., SS14 Km 163.5, I-34149 Basovizza Trieste, Italy}

\author{M. Piccin}
\affiliation{CNR-IOM, SS14 Km 163.5, I-34149 Basovizza Trieste, Italy }
\affiliation{Soitec S.A., F-38190 Bernin, France}

\author{S. Rubini}
\affiliation{CNR-IOM, SS14 Km 163.5, I-34149 Basovizza Trieste, Italy}

\author{F. Martelli}
\affiliation{CNR-IOM, SS14 Km 163.5, I-34149 Basovizza Trieste, Italy }
\affiliation{CNR-IMM via del Fosso del Cavaliere 100 I-00133 - Roma (Italy)}

\date{\today}
\begin{abstract}
We present an EXAFS study of the Mn atomic environment in Mn-doped GaAs nanowires. Mn doping has been obtained either via the diffusion of the Mn used as seed for the nanowire growth or by providing Mn during the growth of Au-induced wires. As a general finding, we observe that Mn forms chemical bonds with As but is not incorporated in a substitutional site. In Mn-induced GaAs wires, Mn is mostly found bonded to As in a rather disordered environment and with a stretched bond length, reminiscent of that exhibited by MnAs phases. In Au-seeded nanowires, along with stretched Mn-As coordination we have found the presence of Mn in a Mn-Au intermetallic compound.
\end{abstract}
%
\pacs{75.50.Pp, 61.72.Vv, 61.10.Ht}
\maketitle
%

\section{Introduction}\label{sec:Introduction}
One of the most promising techniques for the realization of nano-sized electronic devices is the so-called bottom-up approach. With this technique, self-assembly at the atomic level is exploited to produce nanometer-size objects. Examples of the bottom-up approach to the fabrication of nanostructures include semiconductor quantum dots, carbon nanotubes and semiconductor nanowires (NWs). NWs have revealed to be effective in the realization of biochemical sensors \cite{cui_sci_01}, light-emitting diodes \cite{duan_nat_01}, single-electron transistors \cite{thel_apl_03} or solar cells \cite{baxter:053114}. The fabrication of NWs generally occurs via the use of a metal nanoparticle (NP), typically Au, which induces and dictates the growth. The vapor-liquid-solid (VLS) model, initially proposed in the 1960s to produce $\mu$m-sized Si ``whiskers'' \cite{wag_apl_64} and later justified thermodynamically and kinetically \cite{giv_jcg_75}, is generally used to rationalize the growth mechanism. In this model, the metal NP is heated above the eutectic temperature for the metal--semiconductor alloy system and is exposed to a flux of the semiconductor in the vapor phase, producing a liquid nanoparticle droplet. When the concentration of semiconductor atoms increases above saturation, the semiconductor precipitates and a solid semiconductor NW nucleates at the liquid-solid interface. \\
This technique could in principle be used for the production of 1-dimensional magnetic semiconductors. A possible way to realize this class of materials is to disperse a modest concentration (typically 1-5 \%) of the magnetic species (Mn, Fe, Co) in the semiconductor matrix giving rise to the so-called dilute magnetic semiconductors (DMS) \cite{fur_jap_88}. Ideally, the magnetic ions should occupy substitutional sites in the semiconductor lattice and give rise to the crystal magnetic properties. A key issue for these materiasl is their limited Curie Temperature (maximum reported values are 172 K for bulk $p$-type delta-doped GaAs:Mn \cite{naz_prb_03} and 250 K in specially designed heterostructures \cite{nazmul:017201,nazmul:149901}) but if it has been shown that room temperature ferromagnetism could be obtained in wide gap hosts (ZnO or GaN) \cite{die_sci_00}. It is important to note that ferromagnetism can also be observed if ferromagnetic precipitates are present in the material but this does not qualify as a true DMS, of course. On the experimental side, the site of Mn in GaAs has been studied by ion channelling techniques (especially proton-induced x-ray emission, PIXE) and by the extended x-ray absorption fine structure (EXAFS). Using PIXE, the presence of up to 15 \% of Mn in interstitial sites has been detected in GaMnAs alloys \cite{yu_prb_02,yu_nim_04}, possibly in the close vicinity to Mn in substitutional sites. \\
The combination of the two research fields - NWs and DMSs - opens the possibility to envisage 1D spintronics. Attempts to dope semiconductor nanowires with Mn have been reported for GaAs \cite{sadowski-07, kim-09, sadowski-11, Adell-11}, ZnO \cite{ronning:783,chang:4020}, GaN \cite{dee_cpl_03,han:032506,baik:042105}; whereas a number of II-VI and III-V materials are reported having been doped with Mn in a chemical vapour deposition reactor \cite{rad_nl_05}. In all these reports, Mn doping of the semiconductor material is provided by supplying Mn precursors during the growth or by post-growth ion implantation. \\
An alternative method to deposit GaAs NWs using Mn as a growth seed in molecular beam epitaxy (MBE) growth has been reported by our group \cite{mar_nl_06}. It was found that the wires had a wurtzite (WZ) structure and both EXAFS and transport measurements suggested that Mn atoms actually diffuse into the NWs where they act as dopants. A further study \cite{10-jvst-jabeen} was carried out on Mn-induced InAs wires in which morphological investigations showed the effectiveness of this metal in forming wires. In this case the structural study of the Mn environment by EXAFS evidenced the formation of MnAs precipitates.  \\
It is clear from the above that a characterization and a physical understanding of the local structure of Mn in (GaMn)As NWs is of great importance. In the wire growth induced by Mn NPs, two main questions arise: the degree of substitutional incorporation of Mn in the NW lattice and/or whether Mn - defective structures are formed. In this framework, EXAFS can play a decisive role \cite{bosche_2008}. The advantages of this technique in the present context are the chemical selectivity, the high resolution in the determination of the local structure and the applicability both to ordered and disordered atomic arrangements. It is worthwhile to mention some previous EXAFS work on related systems. Soo et al. \cite{soo:2654,soo:2354}, have studied Mn/GaAs digital alloys. They found that in samples deposited at the low temperature of 275 $^\circ$C, Mn substitutes Ga and locally forms a GaMnAs alloy; upon annealing, a noticeable decrease of the first shell coordination number coupled to an increase of the first shell radius was found and the authors suggested that this was due to initial stages of precipitation of a MnAs phase. Later, by combining EXAFS with X-ray magnetic circular dichroism (XMCD) to study the same system, the same group \cite{soo_prb_03} found that a high ferromagnetic alignment of the Mn atoms is linked to a local structure exclusively composed of substitutional Mn in the Ga sites, as in a random (GaMn)As alloy. More recently, some of us \cite{dac_prb_06} have studied a series of Mn $\delta$-doped GaAs samples, deposited at growth temperatures in the range 300 $^\circ$C - 450 $^\circ$C. In low temperature samples, Mn was confirmed to be substitutional to Ga. This study illustrates the ability of EXAFS to detect relatively small fractions of defective site and to measure their local structure quantitatively. \\
In this paper, we report a detailed study using EXAFS of the local structure of Mn in GaAs NWs grown using Mn seeds, following up and extending our previous reports \cite{mar_nl_06, 10-jvst-jabeen}. In the case of Mn-induced GaAs NWs we will report more detailed studies with respect to the preliminary study, with the investigation of samples grown under different conditions. Moreover, particular care has been taken to reduce contributions to the signal coming by post-growth oxidized parts of the NWs. As an interesting comparison we will also present data on Au-induced GaAs NWs that have been doped with Mn during the MBE growth.

\section{Experimental}\label{sec:Experimental} 
\subsection{Sample Preparation and EXAFS measurements}
\begin{figure}[!hbtp]
\includegraphics [width=0.6\linewidth]{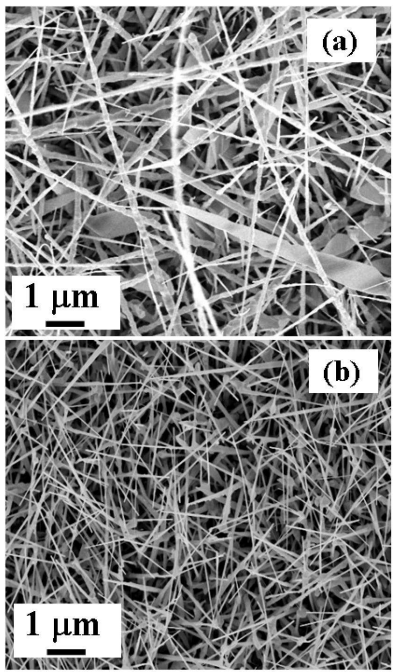}
\caption{\label{fig:micro_1}SEM images of representative samples. (a) E650, Mn-induced GaAs NWs; (b) E697, Au-induced and Mn-doped GaAs NWs.}
\end{figure}

GaAs NWs have been grown by molecular beam epitaxy on SiO$_2$ substrates. Before the NW growth, a thin film of Mn or Au was deposited on the substrates at room temperature in a metallization chamber, connected in ultra-high vacuum with the growth chamber. For all GaAs NW samples, an equivalent two-dimensional growth rate of 1 $\mu$m/h has been used with a V/III beam-equivalent-pressure ratio (BEPR) of 15. Sample growth was terminated by keeping As overpressure on the samples during cooling down. Two types of NWs were studied as a function of growth temperature, nature of the seeding nanoparticle and doping. Samples E650, E651, E672 and E652 were grown using 5 monolayers (ML) of Mn ($\approx$ 1nm) at growth temperatures 540 $^\circ$C, 580 $^\circ$C, 610 $^\circ$C and 620 $^\circ$C, respectively, and the growth was carried out for 30 minutes. Samples E656 and E697 were grown for 30' at 540 $^\circ$C and 580 $^\circ$C, respectively, using 5ML ($\approx$ 1nm) of Au and were doped during the growth with Mn, using an effusion cell. The Mn deposition rates were  $4 \times 10^{12} at/cm^2/s$ in case of sample E656 and roughly one order of magnitude lower in case of sample E697. The details of the preparation of the various samples are summarized in Tab.~\ref{tab:sapre_tab}. The use of growth temperatures higher than those typical for 2D GaAs:Mn layers is dictated by the difficulties found in growing GaAs NWs in the temperature range of 300-400 $^\circ$C.\\
\begin{table}[!hbt]
\caption{\label{tab:sapre_tab} Details of the sample preparation. The equivalent  thickness of the seed (Mn or Au) is in all cases 1 nm.}
\begin{tabular}{|c|c|c|c|c|}
\hline
Sample & Seed Metal & Growth                  & Growth \\
       &            & temperature ($^\circ C$) & time (min) \\
\hline
E650& Mn  & 540 & 30  \\
E651& Mn  & 580 & 30  \\
E672& Mn  & 610 & 30  \\
E652& Mn  & 620 & 30  \\
E656& Au  & 540 & 30  \\
E697& Au  & 580 & 30  \\
\hline
\end{tabular}
\end{table}
The morphology and the structural characterization of similar nanowires can be found in \cite{mar_nl_06} and \cite{piccin-07} for Mn- and Au- induced NWs respectively; a Scanning Electron Microscopy (SEM) picture is presented in Fig.~\ref{fig:micro_1} where the formation of wires and ribbons is evident. Images from samples grown in the given temperature range are very similar thus suggesting a smooth dependence of the growth rate on temperature.\\
EXAFS data at the Mn-K edge (6539 eV) have been collected at the GILDA-CRG beamline operative at the European Synchrotron Radiation Facility \cite{dac_enl_98}. The monochromator was equipped with a pair of Si(111) crystals and was run in dynamically focusing mode \cite{pas_jsr_96}. Harmonic rejection was achieved by using a pair of Pd-coated mirrors with an energy cutoff of 19.5 keV. The intensity of the incoming beam was measured with a $N_2$ filled ion chamber whereas the fluorescence signal from the Mn-K$_\alpha$ line was measured with a 13 elements Hyper Purity Germanium detector. Measurements were carried out at T=10 K using a liquid He cryostat in order to minimize the damping of the EXAFS signal originated from thermal atomic vibrations. Two to four spectra per sample were collected in order to improve the Signal-to-Noise ratio. Prior each EXAFS measurement, the samples were chemically etched in HF for 3 seconds to remove the oxide layer, formed because of the exposure to air after the growth, and then immediately loaded into the vacuum chamber. \\
%
\subsection{Extended X-ray Absorption Fine Structure data}
EXAFS data were extracted from the raw absorption spectra using the ATHENA code \cite{rav_jsr_05} and were fitted in R space with the ARTEMIS code \cite{rav_jsr_05} using theoretically calculated backscattering paths. The theoretical paths were calculated with the FEFF8.4 code \cite{ank_prb_98} using model clusters that will be described in the following sections. \\
\subsubsection{Mn-induced GaAs nanowires} 
\begin{figure}
\includegraphics [width=0.6\linewidth, angle=-90]{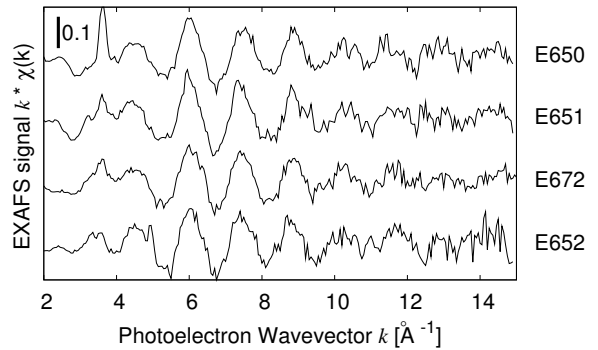}
\caption{\label{fig:exa_mncat} EXAFS data of the samples of Mn-induced GaAs (E650, E651, E672, and E652) nanowires obtained using Mn as seed. The bar marks the scale of the vertical axis.}
\end{figure}

\begin{figure}
\includegraphics [width=0.6\linewidth, angle=-90]{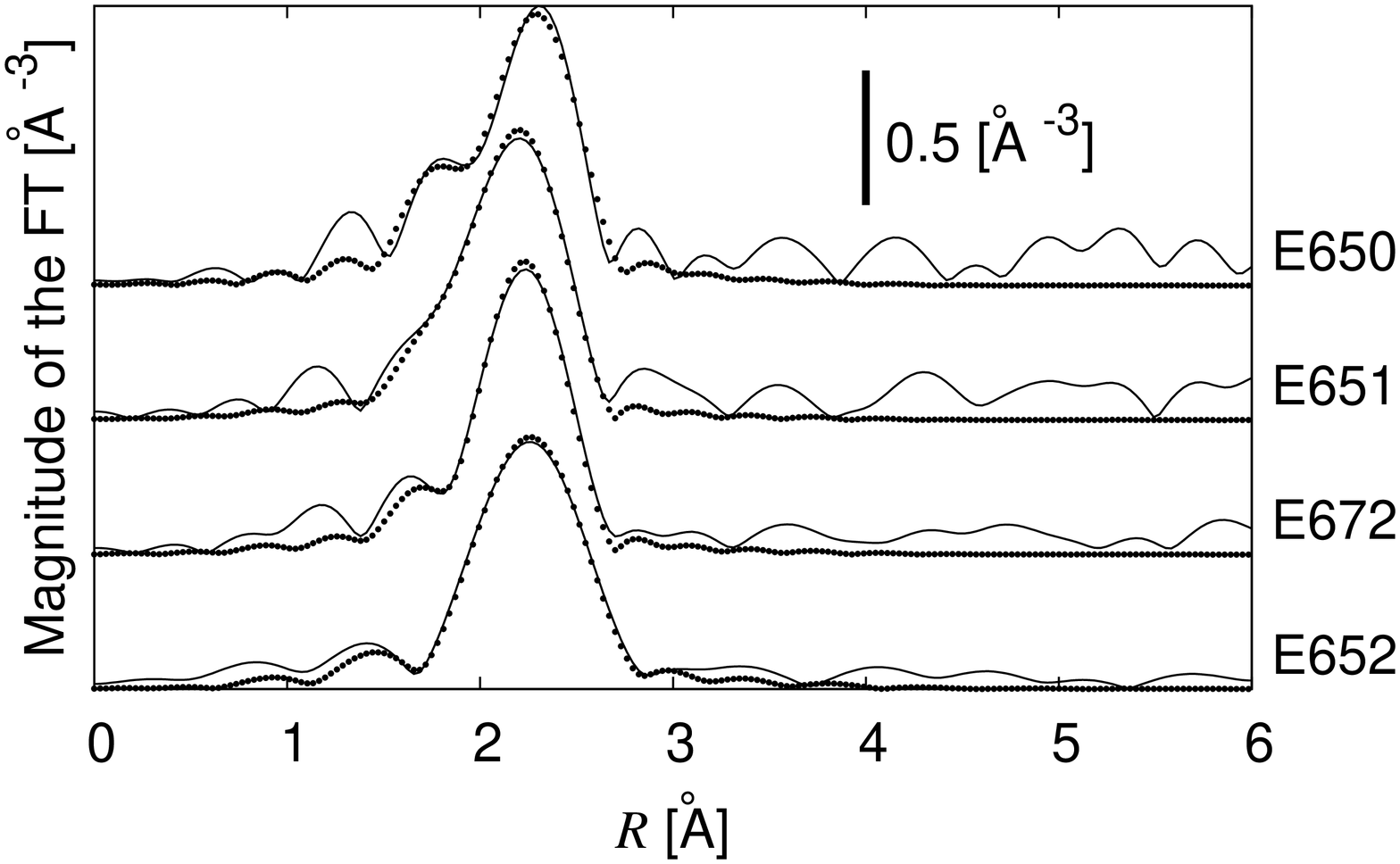}
\caption{\label{fig:fou_mncat} Fourier Transforms of the data presented in Fig.~\ref{fig:exa_mncat} (lines) and best fit curves (dots). Data were transformed in the interval k = [3.2-12.9] \AA \thinspace with a k$^2$ weight and fitted in R space in the interval 1.5-3.0 \AA. The bar marks the scale of the vertical axis.}
\end{figure}

The background subtracted EXAFS data of Mn-induced GaAs NWs are presented in Fig.~\ref{fig:exa_mncat} whereas the related Fourier Transforms (FTs) are presented in Fig.~\ref{fig:fou_mncat}. The results of the quantitative analysis are presented in Tab.~\ref{tab:mncat_tab}. In the EXAFS spectra of the GaAs samples only a single-frequency oscillation is visible and in the related FT there is only a dominating structured peak at $R \approx 2.2$ \AA. This means that only coordinations with a first neighbor can be detected and a considerable structural disorder prevents the detection of higher coordination shells. Some GaAs samples present a shoulder at lower $R$ values in Fig.~\ref{fig:fou_mncat}) that was interpreted as a contribution from a surface oxide phase.
In the analysis of these samples we considered a model consisting in Mn-As bonds accounting for Mn in a Ga-substitutional site in the GaAs matrix ($Ga_{Mn}$, see \cite{dac_prb_06}) plus an oxide phase as already found in \cite{mar_nl_06}. In our previous report the oxide phase was much more intense because no etching of the samples was carried out prior EXAFS measurements. Contributions from $MnO_{x}$ phases are  reduced by this process althought not completely eliminated. An issue about data interpretation could be the contribution of Mn-Mn bonds coming from $\alpha$-Mn in the wire tips, keeping in mind the similar backscattering amplitude between As and Mn. However this can ruled out because the fit of metallic Mn yields an average $R_{MnMn}=2.69$ \AA \thinspace, considerably longer respect to the value in the wires. A futher argument on this aspect will be given in the Discussion section. \\
%
%
%
%
\begin{table}
\caption{\label{tab:mncat_tab} Results of the quantitative EXAFS analysis on the Mn-induced GaAs wires. For the Mn-O bond a length of about 2.16(3) \AA \thinspace was found with a coordination number decreasing from 0.6(4) to 0.0(4) for increasing growth temperature. From the sum of the ionic radii available in literature( $R_{MnO}=2.22$  \AA \thinspace \cite{69-ac-shannon} ) this can be interpreted as $Mn^{2+}$ in a VI coordinated environment.}
\begin{tabular}{|c|c|c|c|c|c|c|}
\hline
Sample &  $N_{As}$ & $R_{As}$ & $\sigma^2_{As}$  \\
       &           &  \AA     &*10$^{-3}$\AA$^2$ \\
\hline
E650   & 2.0(3)    & 2.59(1)   & 5(1)            \\
E651   & 3.2(3)    & 2.56(1)   & 8(1)            \\
E672   & 2.4(3)    & 2.57(1)   & 6(1)            \\
E652   & 3.2(3)    & 2.60(1)   & 9(1)            \\
\hline
\end{tabular}
\end{table}
%

\subsubsection{Au-induced Mn-doped GaAs nanowires}
\begin{figure}
\includegraphics [width=0.6\linewidth, angle=-90]{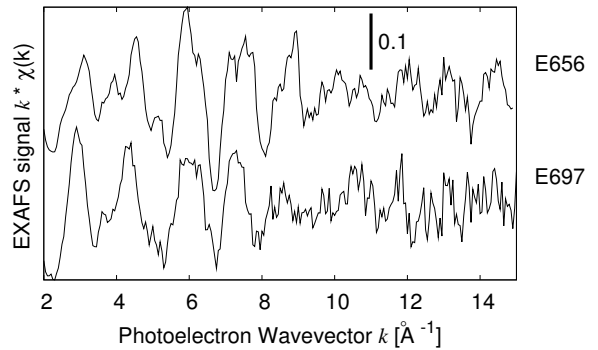}
\caption{\label{fig:exa_aucat} EXAFS data of the samples of Au-induced and Mn-doped GaAs nanowires. The bar marks the scale of the vertical axis.}
\end{figure}

\begin{figure}
\includegraphics [width=0.6\linewidth, angle=-90]{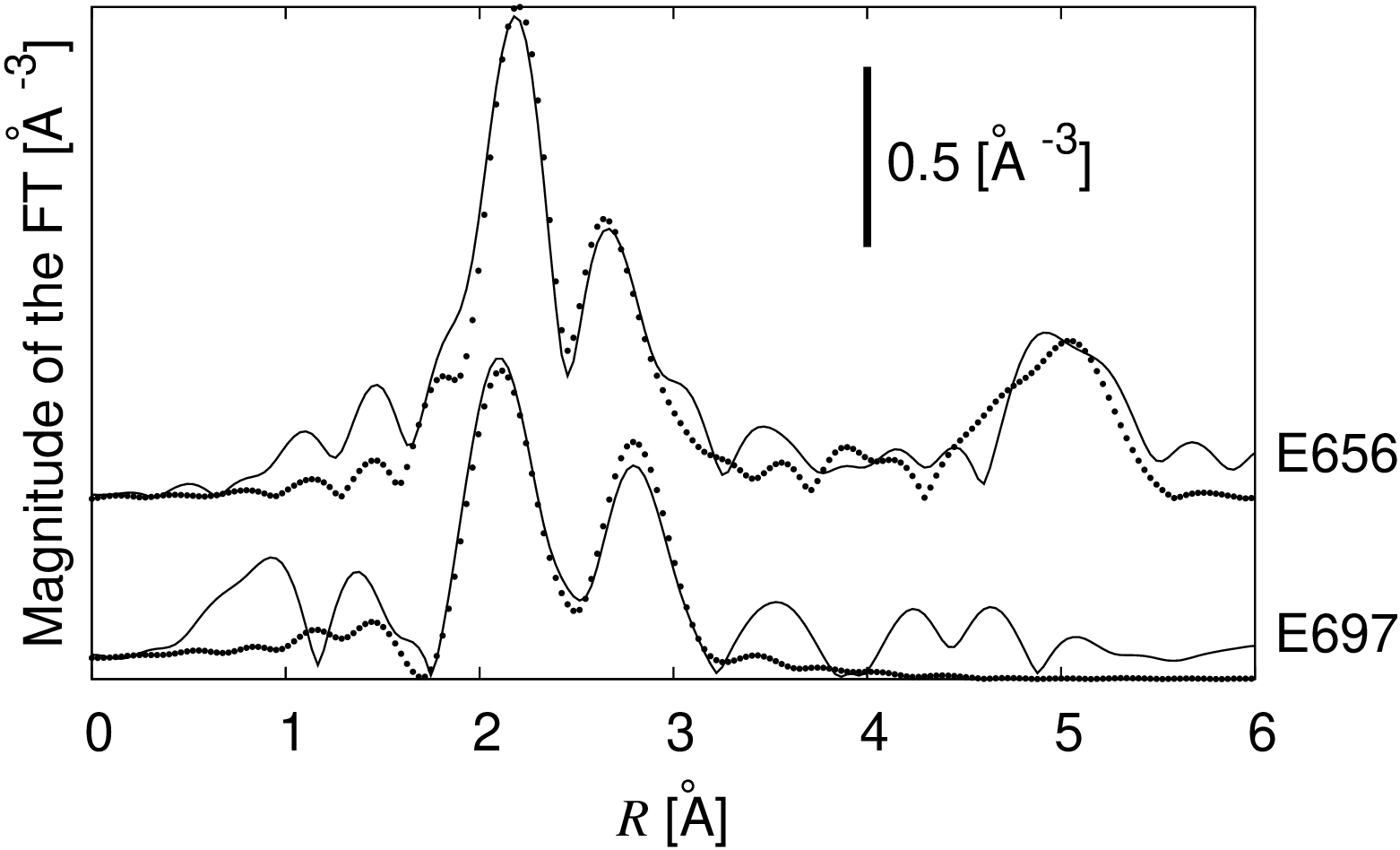}
\caption{\label{fig:fou_aucat} Fourier Transforms of the data presented in Fig.~\ref{fig:exa_aucat} (lines) and best fit curves (dots).  Data were transformed in the interval k = [3.0-14.4] \AA \thinspace with a k$^2$ weight and fitted in R space in the interval 1.2-5.5 \AA. The bar marks the scale of the vertical axis.}
\end{figure}

The EXAFS signals of the samples belonging to this class are shown in Fig.\ref{fig:exa_aucat} whereas the related FT are presented in Fig.\ref{fig:fou_aucat}. Tab.~\ref{tab:aucat_tab} reports the results of the quantitative analysis. An oscillating signal, still present at high k values (i.e. above k=10 \AA $^{-1}$), is particularly well evident in sample E656 (Fig.\ref{fig:exa_aucat}) revealing the presence of a heavy backscatterer around Mn. In the FT a double peak is present in both samples in the region 2-3 \AA \thinspace and an additional broad peak is present at about 5 \AA \thinspace in the E656 sample. In this case, we tested two models consisting of Mn substitutional in GaAs plus i) Mn in a tetrahedral interstitial site or ii) a cubic MnAu (Space group P M -3 M , lattice parameter a=3.2220 \cite{morr_56}) phase. The choice of model i) was suggested by the fact that the interstitial site in the GaAs structure could in principle give rise to a double peak in the first shell, as evidenced in previous studies \cite{dac_prb_06}. However, model ii) best fitted the data; notice that the broad peak in the region around 5 \AA \thinspace is well reproduced by the collinear Mn-Au-Mn paths present in the high R part of the cubic structure whereas it is remains completely unexplained by model i). In sample E697 the peak at 5.5 \AA \thinspace  is not present, while the other features are very similar to those exhibited by sample E656. The local environment in this sample is similar, with a greater disorder (both configurational and chemical) than in sample E697. \\
%
%
%
%
\begin{table}
\caption{\label{tab:aucat_tab} Results of the quantitative EXAFS analysis on the Au-induced GaAs wires doped with Mn.}
\begin{tabular}{|c|c|c|c|c|c|c|}
\hline
Sample &  $N_{As}$ & $R_{As}$ & $\sigma^2_{As}$ & $N_{Au}$ & $R_{Au}$ & $\sigma^2_{Au}$  \\
       &           &  \AA     &*10$^{-3}$\AA$^2$&          &   \AA    &*10$^{-3}$\AA$^2$ \\
\hline
E656& 3.5(8) & 2.56(1) & 8(2) & 0.8(6) & 2.75(1) &0.5 \\
E697& 2.2(7) & 2.58(1) & 9(2) & 3.1(5) & 2.82(1) &5 \\
MnAu \cite{morr_56}&   -    &  -      &  -   & 8     & 2.79    & - \\
\hline
\end{tabular}
\end{table}
%
\section{Discussion}\label{sec:Discussion} 
We will first discuss whether the Mn EXAFS signal, is originated from the Mn in substitutional sites or from different locations in the matrix. 
As a first step, the spacial origin of the Mn-related signal will be discussed as it could either come from Mn within the wires or in the metallic droplets located at the wire tip. It is possible to estimate the ratio of the contributions of Mn in the tips and in the wires by a simple argument. The total amount of Mn deposited on the substrate is 1nm equivalent for a total volume or $10^6$ nm$^3$ for region of 1 $\mu m ^2$. In the same region SEM images exhibit droplets on the wires tip with a radius of 15nm with a consequent volume of $\approx 14000$ nm$^3 $ each. Considering that about 10 droplets/$\mu m ^2$ are present on the images the amount of Mn in this phase is 140*10$^3$ nm$^3$ corresponding to  about 14 \% of the total Mn deposited on the considered region. This is too low to be seen by EXAFS and the signal from Mn comes predominantly from the metal contained in the wires.\\
From the EXAFS quantitative analysis it is clear that Mn in GaAs wires does not occupy substitutional sites. This appears to be true independently of the growth process, using Mn as the growth seed or doping the wires with Mn during the growth of Au induced NWs. First of all the first shell bond length is $R_{MnAs} = 2.59$ \AA \thinspace, which is appreciably longer than that reported for the substitutional site in GaAs ($R_{MnAs} = 2.50$ \AA \thinspace). The particular crystal structure of the NWs (Wurtzite) cannot explain this difference. Indeed, in the commonly encountered ZB structure, the elongation of the Mn-As bond, respect to the Ga-As bond, is only about 2 \%. In the WZ structure a similar modification of the interatomic distance is to be expected because the chemical interaction between Mn and As is the same in the two cases. The length of the GaAs bond in WZ GaAs, using the crystallographic determination of Ref.~\cite{yeh_prb_92} ( Space Group P63MC, lattice parameters a=3.912 \AA \thinspace, c=6.441\AA \thinspace, u=0.374), results to be $R_{GaAs} = 2.399$ \AA \thinspace.  Supposing the same ratio for the $R_{MnAs} / R_{GaAs}$ distances as in ZB, in WZ we should have $R_{MnAs} = 2.45$ \AA \thinspace i.e. a value well below the observed data. On the other hand, the experimental value is very close to the first shell distance in hexagonal MnAs ($R_{MnAs} =2.58$ \AA \thinspace). Similar values have been also found in previous studies on Mn in GaAs \cite{mar_nl_06} and InAs \cite{10-jvst-jabeen} NW obtained by using Mn as seed. In bulk samples this value of bond length, associated to a coordination number Mn-As of about 2.5, was observed in MnAs-GaAs digital alloys annealed at 550 $^\circ C$ \cite{soo:2354} and interpreted as due to the formation of a precursor phase for MnAs particles. Similar results on the Mn-As bond length are reported on (Ga Mn)As alloys annealed at 600 $^\circ C$ \cite{07-jpcm-dem}. The samples of the present study are definitely different from those presented in Ref.~\cite{dac_prb_06} and this is reasonably due to the higher temperature used in their preparation (see Tab.~\ref{tab:sapre_tab}). The low value of the coordination number (about 3 instead of 6) suggests that, rather than creating extended crystals of the hexagonal phase, Mn forms small precursors of this structure. A noticeable level of structural disorder prevents the observation of the higher coordination shells, in contrast to what previously observed in ZB-GaAs \cite{dac_prb_06}. \\
Finally, the results for Au-induced and Mn-doped samples are discussed. Similarly to previous cases, the Mn-As bond length is longer than expected for a substitutional impurity. Moreover, Mn-Au bonds have been detected, indicating the formation of an intermetallic compound with evidence of chemical ordering via the presence of collinear paths Mn-Au-Mn. The degree of ordering of the alloy depends on the growth temperature, sample grown at 580 $^\circ$C exhibiting a marked chemical ordering. This affinity between Mn and Au suggests that Mn participates to the wire growth process creating a supersaturated alloy with Au in the head similarly to what happens to Ga.
 For the considerations given above, it is reasonable that islands of MnAu alloy remain in the NW during the growth. A similar effect has been reported in the preparation of ternary $In_xGa_{1-x}As$ NWs \cite{jab_apl_08}. This might indicate a more general behavior, that is the participation of impurity atoms to the growth process of the nanowires. \\
Since a pure substitutional Mn site has not been observed in the samples we have studied, it appears that growth parameters different from the ones we have used must be envisaged to dope III-V nanowires with Mn at a high efficiency. The origin of the defective incorporation of Mn is most probably the relatively high growth temperature which is needed to produce NWs themselves; we recall indeed that only at lower temperatures Mn is found in substitutional sites, for example in  $\delta$-doped Mn:GaAs structures \cite{dac_prb_06}. The presence of a minor amount of substitutional Mn can not be ruled out from these data considering that the detection limit for the amount of such a secondary phase against the dominating one can be estimated to be around 5 \%. An important step in future works will be the attainment of the NW growth at lower temperatures that will presumably raise the fraction of Mn in substitutional sites.
\section{Conclusion}\label{sec:Conclusion}
The analysis of the atomic environment of Mn impurities incorporated in GaAs nanowires has revealed that Mn forms chemical bonds with As with a bond distance of 2.56-2.58 \AA \thinspace. This value is longer than expected for a substitutional site and is probably due to the occupation of defect sites that are the seeds for the formation of hexagonal MnAs precipitates. This phenomenon is observed both in Mn- or Au- induced nanowires where the doping procedure is radically different. In the last case the presence of MnAu intermetallic phase has been also revealed by the presence of Mn-Au intermetallic bonds.
\section{Acknowledgements}\label{sec:Acknowledgements} 
The authors are indebted to H. Pais for technical assistance on the GILDA beamline. GILDA is a project jointly financed by CNR and INFN.
%
\bibliographystyle{apsrev}
\bibliography{nwrs_v17}

\end{document}